\documentclass{article}
\usepackage[utf8]{inputenc}
\usepackage{natbib}
\usepackage{booktabs}
\usepackage{adjustbox}
\usepackage[a4paper, total={6in, 8in}]{geometry}
\usepackage{graphicx}
\usepackage{amsmath} 
\usepackage{rotating,tabularx,booktabs,siunitx,caption}
\usepackage{setspace}
\usepackage{booktabs}
\usepackage{xltabular}  
\usepackage{eurosym}
\usepackage{siunitx}
\usepackage{comment}
\usepackage{xtab}
\newcommand{\quotestart}{``}
\newcommand{\quoteend}{''}
\usepackage{url}

\newtheorem{definition}{Définition}[section] 

\setcitestyle{authoryear,open={(},close={)}} 

\title{Crosswashing in Sustainable Investing: Unveiling Strategic Practices Impacting ESG Scores}
\date{\today}

\begin{document}

\maketitle

{\centering

Bertrand Kian Hassani \textsuperscript{1, 3, 4, 5},
Yacoub Bahini \textsuperscript{2}

\vspace{4mm}
\textsuperscript{1,2}\textit{QUANT AI Lab, C. de Arturo Soria, 122, 28043 Madrid, Spain}
\textsuperscript{3}\textit{Department of Computer Science, University College London, Gower St, London WC1E 6EA, UK}
\textsuperscript{4}\textit{Centre d'Économie de la Sorbonne (CES), Maison des Sciences Économiques (MSE), Université Paris1 Panthéon Sorbonne, 106-112 Boulevard de l’Hôpital, 75013 Paris, France}
\textsuperscript{5}\textit{Institut Louis Bachelier,  Palais Brongniart, 28 Pl. de la Bourse, 75002 Paris}

\vspace{4mm}

\textsuperscript{1}\textit{ bertrand.hassani@quant.global}

\textsuperscript{2}\textit{yacoub.bebekar@quant.global}

\begin{abstract}
This paper introduces and defines a novel concept in sustainable investing, termed \textit{crosswashing},  and explore its impact on ESG (Environmental, Social, and Governance) ratings through quantitative analysis using a Multi-Criteria Decision Making (MCDM) model. The study emphasises that this specific form of greenwashing is not currently considered in existing ESG assessments, potentially leading to an inflated perception of corporate ethical practices. Unlike traditional greenwashing, crosswashing involves companies strategically investing in sustainable activities to boost Environmental, Social, and Governance (ESG) scores while preserving non-sustainable core operations.  By unveiling the nuances of crosswashing, the research contributes to a more nuanced understanding of sustainable investing, offering insights for improved evaluation and regulation of corporate environmental and ethical responsibilities.
\end{abstract}
}
~

\section{Introduction}

In recent times, there has been a shift from mere adherence to legal norms to fully embracing ethical principles. This transformation is driven by the growing significance of reputation and the associated risks to one's public image (\cite{su8111196}) to appear compliant with \textit{sustainable development}, which is  defined as “development
that meets the needs of the present without compromising
the ability of future generations to meet their own
needs” (\cite{wced1987world}).\\

\noindent Over the past decade, sustainable investments have witnessed a remarkable surge in popularity and growth. This surge can be attributed to a combination of factors, including the increasing awareness of environmental and social issues (\cite{rajapaksa2018pro}, \cite{uralovich2023primary}), changing consumer preferences (\cite{oroian2017consumers}, \cite{bask2013consumer}, \cite{rajapaksa2018pro}, \cite{uralovich2023primary}), and a greater emphasis on corporate social responsibility(\cite{moon2007contribution}, \cite{herrmann2004corporate}, \cite{weber2014corporate}). An increasing number of companies are now promoting the environmental friendliness of their products and practices, aiming to capitalise on the growing opportunities presented by expanding green markets (\cite{doi:10.1525/cmr.2011.54.1.64}). \quotestart As there is an increase in green markets, it is followed by the phenomenon of \textit{greenwashing}" (\cite{de2020concepts}, \cite{majlath2017effect}).  The phenomenon of \textit{greenwashing}\footnote{\cite{de2020concepts} gives some examples of greenwashing} is defined as \quotestart the intersection of two firm behaviours: poor
environmental performance and positive communication
about environmental performance" (\cite{doi:10.1525/cmr.2011.54.1.64}).
\quotestart \textit{Greenwashing} is the act of misleading consumers regarding the environmental practices 
of a company (firm-level) or the environmental benefits of a product or service
(product/service-level)\quoteend (\cite{de2020concepts}).\\

\noindent \quotestart The current environment of failures of corporate responsibility are not only failures of legal compliance, but more fundamentally failures to do the right (ethical) thing" (\cite{arjoon2005corporate}). While there have been some explanations for firm greenwashing, there is a lack of a comprehensive analysis of the factors driving it. Consequently, there are limited tools available for managers or policymakers to effectively address and mitigate greenwashing (\cite{doi:10.1525/cmr.2011.54.1.64}). In this context, the path to doing the right thing in sustainability appears less evident and demands further exploration. The current resulting regulatory environment might not be comprehensive enough..

\noindent \textit{Sustainable investing} is typically evaluated based on environmental, social, and governance (ESG) scores (\cite{gibson2022responsible}, \cite{liang2022responsible}, \cite{dumitrescu2022defining}). \quotestart Hence, asset managers that claim
to invest according to ESG principles but invest in firms with low ESG scores are often accused of
greenwashing" (\cite{dumitrescu2022defining}). Nevertheless, in the absence of a clear definition of sustainable investing, assessing the extent of greenwashing becomes challenging. Understanding the factors that facilitate or hinder this practice is also impeded (\cite{dumitrescu2022defining}).\\

\noindent In this context, the paper endeavours to contribute  to the literature by defining a specific  aspect of \textit{sustainable investing}. In particular, our focus lies on a company's investments in \textit{sustainable activities} that are directly or indirectly associated with its their primary operations, especially when the \textit{main activity} is not inherently sustainable.  Indeed, companies' investments in sustainable activities enhance their ESG scores without necessarily diminishing the negative impact of their core business activities.  \cite{raghunandan2022esg} present that \quotestart ESG scores are correlated with the quantity of voluntary ESG-related disclosures but not with firms’ compliance records or actual levels of carbon emissions". This statement summarises this paper's underlying motivation.
This practice should be considered as a distinct category within greenwashing, and it is refered to as \textit{crosswashing}. 
To the best of our knowledge, crosswashing is not regarded as a type of greenwashing practice because it involves strategically creating an image of moral responsibility, providing protection against significant social risks, all while ensuring compliance with existing regulations. Thus, companies' ESG scores do not consider this dimension of greenwashing, and as a result, ESG scores could be overestimated, as we will explore further in this paper. Moreover, the tendencies of GHG emissions, for example, continue to grow, necessitating a revision of current ESG scores. \quotestart Countries are expected to emit a total 36.8 billion metric tons of CO2 from fossil fuels in 2023, a 1.1\% increase from last year, the report by scientists from more than 90 institutions including the University of Exeter concludede" (\cite{reuters2023co2}).

\section{Greenwashing: A litterature overview}
\label{Greenwashing_overview}
\subsection{Greenwashing typology in literature}
\label{Greenwashingtypologyinliterature}

\cite{doi:10.1177/1086026615575332}  note the absence of a rigid definition for greenwashing, attributing it to its multifaceted nature. In 1986, environmentalist Jay Westervelt introduced the term \quotestart green-washing" through an essay that delved into the hospitality industry. Since then, numerous academic definitions of the concept have emerged (\cite{GUO2018127}, \cite{doi:10.1177/1086026615575332}).\\

\noindent Following \cite{doi:10.1177/1086026615575332}, we can argue that the definition of \textit{greenwashing} varies depending on the perspective from which it is considered. The authors considers two main approaches to address this ambiguity  based on the quality of disclosure, the first of which is  \textit{greenwashing as selective disclosure}. Subsequent to this viewpoint, the prevailing literature characterises the phenomenon through a company’s two primary behaviours: withholding the revelation of adverse details linked to its environmental performance and highlighting positive information about its environmental achievements. This dual behaviour is commonly referred to as selective disclosure\footnote{Consult the work of the authors \cite{doi:10.1177/1086026615575332} for diverse definitions found in the literature.}.\\ 

\noindent The second approach is  \textit{greenwashing as decoupling}, which is  based on decoupling behaviour. According to \cite{doi:10.1177/1086026615575332}, 
the different definitions following this approach consider that greenwashing involves the employment of symbolic environmental actions without genuine impact or failing to fulfill environmental commitments. This strategic use of symbols aims to alleviate external pressures, uncertainties, and conflicts with stakeholders, ultimately maintaining corporate legitimacy.\\

\noindent In the context of greenwashing, these  two approaches are linked with the \textit{signalling and corporate legitimacy theory\footnote{See \cite{Legitimacy} for more details on legitimacy theory.}} in the context of greenwashing. Authors (\cite{doi:10.1177/1086026615575332})  distinguishing three types of corporate legitimacy: cognitive legitimacy (perceived importance among constituents), 
pragmatic legitimacy (benefits for constituents), and moral legitimacy (positive evaluation of green practices) (see \cite{Legitimacy} and \cite{https://doi.org/10.1002/bse.1912}). \cite{Legitimacy} give a very detailed analysis of the legitimacy concept. We have kept on only these three legitimacy types based on the studies of \cite{doi:10.1525/cmr.2011.54.1.64} and \cite{https://doi.org/10.1002/bse.1912}. \quotestart Following the legitimacy literature (\cite{oliver1991strategic}; \cite{suchman1995managing}; \cite{walker2012harm}), and in particular the strategic approach to pragmatic legitimacy as conceptualised by \cite{scherer2013managing}, we argue that green and social marketing, regardless of its degree of falsehood, is a strategy to obtain pragmatic legitimacy. The risk here is that companies can be accused of greenwashing" (\cite{https://doi.org/10.1002/bse.1912}). So \cite{https://doi.org/10.1002/bse.1912} consider that  greenwashing falls within the scope of pragmatic legitimacy since it is “the result of self-interested
calculations of the organisation’s key stakeholders, and it is based on stakeholder’s perceptions of their personal benefit deriving from corporate activities and communication"  (\cite{https://doi.org/10.1002/bse.1912}).
On the other hand, in situations where companies fall short of their environmental targets, according to the insights provided by \cite{BioTechnologyAnIndianJournal}, decoupling behaviours may lead to a reduction in the three legitimacy types. Notably cognitive legitimacy, moral legitimacy, and pragmatic legitimacy."

\subsection{Characteristics of greenwashing}
\label{Characteristics_of_Green_washing}

In the literature, greenwashing characteristics are classified in various ways, and we can distill them into three primary classifications.  First, the \textbf{claim greenwashing} concept, \quotestart which uses textual arguments that explicitly or implicitly refer to the ecological benefits of a product or service to create a misleading environmental claim" (\cite{de2020concepts}). This claim can be expressed as levels of  \textbf{deceptive claims}, which \quotestart could be summed up as lying, lying by omission or lying through lack of clarity" (\cite{de2020concepts}). It can can also be expressed as a\textbf{claimn type} that \quotestart involves five typological categories: (a) product orientation, claims centring on the ecological attribute of a product; (b) process orientation, claims centring on the ecological high performance of a production process technique, and/ or an ecological disposal method; (c) image orientation, claims centring on enhancing the eco-friendly image of an organisation, like claims that associates an organisation with an environmental cause or activity which there is elevated public support; (d) environmental fact, claims that involves an independent statement that is ostensibly factual in nature from an organisation about  the environment at large, or its condition; and (e)
combination, claims having two or more of the categories above [2, 47]" (\cite{de2020concepts}). The analysis of the greenwashing claim in the literature concentrates on the level of products and services (\cite{de2020concepts}).\\

\noindent Second is the classification called \textit{the thirteen sins of greenwashing}: (1) hidden trade-off, (2) no proof, (3) vagueness, (4) worshipping false labels, (5) irrelevance, (6) lesser of two evils, (7) fibbing, (8) false hopes, (9) fearmongering, (10) broken promises, (11) injustice, (12) hazardous consequences, and (13) profits over people and the
environment. \cite{de2020concepts} cite the seven sins of \cite{terrachoice2010} completing them from \cite{scanlan2017framing}'s version. In the following sections, we will briefly elaborate on  certain aspects of the sins of greenwashing when addressing \textit{crosswashing}.\\

\noindent Third, the typology we call \textit{firm-greenwashing transgressions}. It includes five firm-level greenwashing  outlined in \cite{contreras2017fuzzy}, drawing from the work of \cite{bruno1992greenpeace} and \cite{berrone2016green}.
Here are their details: 
\begin{itemize}
    \item{Dirty business:} Belonging to an inherently unsustainable business, but promoting sustainable 
practices or products that are not representative either of the business or the society. 
    \item{Ad bluster:} Diverting attention from sustainable issues through the use of advertising. It is 
used to exaggerate achievements or present alternative programs that are not 
related to the main sustainability concern.
    \item{Political spin:} Influencing regulators or governments in order to obtain benefits that affect 
sustainability. It is common to notice that these spins are “justified” due to the 
companies’ character of large tax payers or employers.
    \item{It’s the law, stupid!:} Proclaiming sustainability accomplishments or commitments that are already 
required by existing laws or regulations.

    \item{Fuzzy reporting:} Taking advantage of sustainability reports and their nature of one-way 
communication channel in order to twist the truth or project a positive image in 
terms of CSR corporate practices
\end{itemize}

\noindent Fourth, \textbf{executional greenwashing} with the following definition by  \cite{parguel2015can}: \quotestart This strategy of greenwashing does not use any type of claim that was described before, but it suggests nature-evoking elements such as images using colours (e.g., green, blue) or sounds (e.g., sea, birds). Backgrounds representing natural landscapes (e.g., mountains, forests, oceans) or pictures of endangered animal species (e.g., pandas, dolphins) or renewable sources of energy (e.g., wind, waterfalls) are examples of executional nature-evoking
elements" (\cite{de2020concepts}).\\

\noindent  Other classifications and strategies of greenwashing are mentioned in the literature (see \cite{de2020concepts}, \cite{parguel2015can}); however, we specifically reference these four forms as they encompass all others.

\subsection{Greenwashing drivers}
\label{GreenwashingDrivers}
Individual psychological drivers: The drivers at the individual level encompass factors, including  \textit{narrow decision framing}, \textit{hyperbolic intertemporal discounting}, and \textit{optimistic bias} (\cite{doi:10.1525/cmr.2011.54.1.64}). The influences of uncertainty and limited or imperfect information directly impact these individual components, with the current regulatory environment contributing to these factors (\textit{ibid.}). \quotestart Narrow decision framing" involves making decisions in silos, where employees or managers may choose to positively communicate about a firm's \quotestart greenness" without considering the actual requirements for future implementation (\textit{ibid.}). \quotestart Hyperbolic intertemporal discounting", on the other hand, is the inclination of individuals to exhibit high impatience over short time horizons and a low level of patience over longer time horizons (\textit{ibid.}). \quotestart Hyperbolic
discounting generates what is often referred to as dynamic inconsistency, or preference reversals" (\cite{doi:10.1525/cmr.2011.54.1.64}). Finally, optimistic bias characterises the inclination of individuals to overestimate the probability of positive events while underestimating the likelihood of less favourable events (\textit{ibid.}). Thus those characteristics deviates the firms’ deciders short term acts
from long term decisions and objectives.\\

\noindent \textit{Organisational level of drivers}: The \textit{organisational level of drivers} consists of \textit{firm characteristics}, \textit{incentive structure and
ethical climate}, \textit{organisational inertia}, and \textit{effectiveness of intra-firm communication} (\textit{ibid.}). \textit{Firm characteristics}, according to \cite{doi:10.1525/cmr.2011.54.1.64}, influence a company's likelihood of being targeted for greenwashing investigations by environmentalists and NGOs, with larger firms or those with negative environmental impacts facing increased scrutiny and potential legal action. \\
\noindent \textit{Incentive structure and ethical climate} are two factors that contribute to unethical behavior, potentially leading to greenwashing, as they involve bypassing moral and ethical standards in favor of decisions driven by an egoistic perspective (\textit{ibid.}). \textit{Organisational inertia}
 is the phenomenon wherein statements or intentions regarding the implementation of environmentally friendly changes within a company are discussed by management but not effectively translated into action, thereby constituting a form of greenwashing (\textit{ibid.}). The \textit{effectiveness of intra-firm communication} is defined as a driver, stemming from the company's challenge in disseminating knowledge internally. Firms with ineffective communication between departments or sections are more likely to greenwash (\textit{ibid.}).\\

\noindent \textit{Market external drivers:} The \textit{market external drivers} encompass three key factors: \textit{consumer demand}, \textit{investor demand}, and \textit{competitive pressure}. Firms are compelled to embellish their images with false information due to the pressure from consumers and investors, as well as the desire to appear greener than their competitors (\textit{ibid.}).\\

\begin{figure}[h]
    \centering
    \includegraphics[width=10cm]{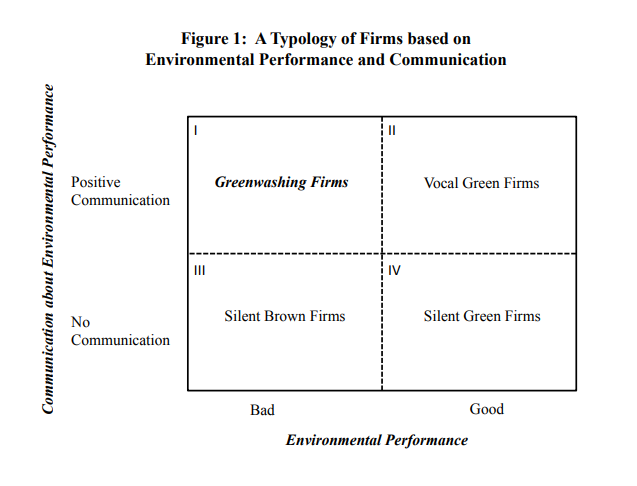}
    \caption{A Typology of Firms based on 
Environmental Performance and Communication . (\cite{doi:10.1525/cmr.2011.54.1.64}).}
    \label{DelmasandBurbano1}
\end{figure}

\begin{figure}[h]
    \centering
    \includegraphics[width=14cm]{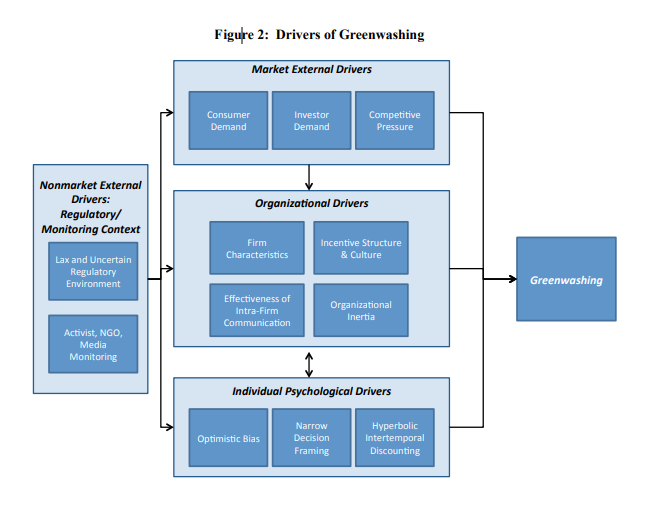}
    \caption{Drivers of Greenwashing. Source (\cite{doi:10.1525/cmr.2011.54.1.64}).}
    \label{DelmasandBurbano2}
\end{figure}

\begin{figure}[h]
    \centering
    \includegraphics[width=12cm]{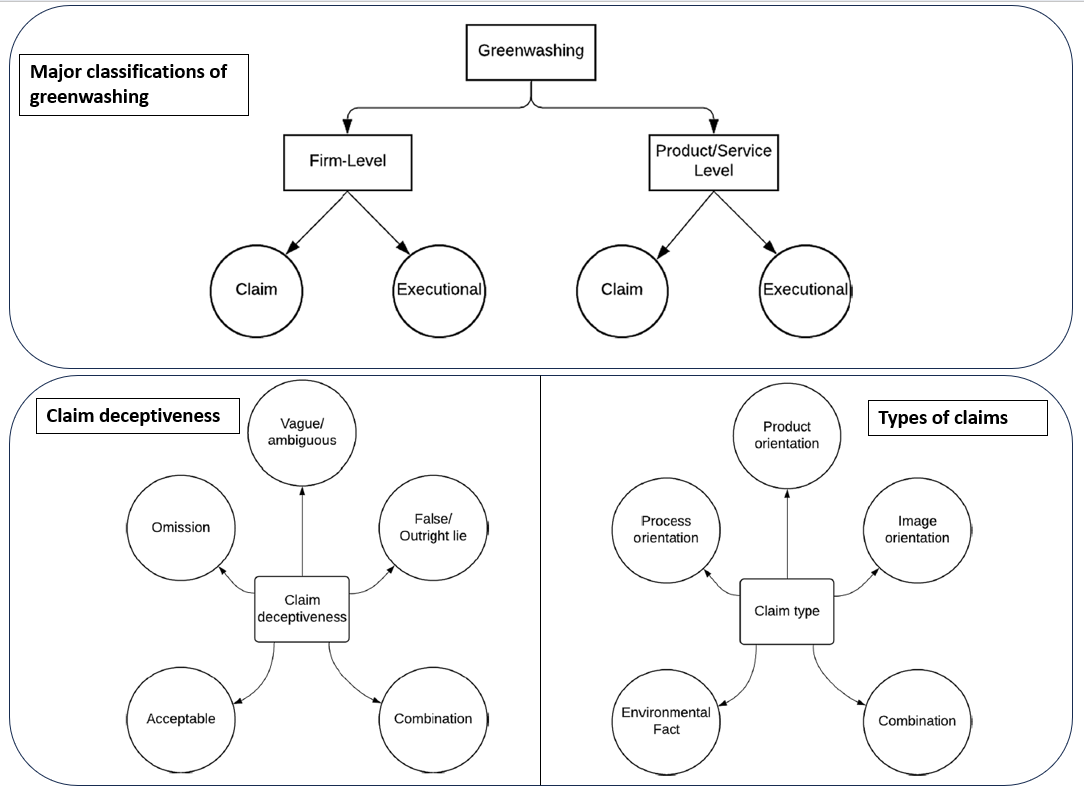}
    \caption{Drivers of Greenwashing. Source (\cite{de2020concepts}).}
    \label{DelmasandBurbano3}
\end{figure}

\noindent \textit{Regulatory context:} The regulatory context is a critical direct driver of greenwashing due to the limited punitive consequences of greenwashing. Furthermore, the regulatory context influences the external market, organisational and individual-level drivers of greenwashing\footnote{See Figures \ref{DelmasandBurbano1}, \ref{DelmasandBurbano2}, \ref{DelmasandBurbano3}.}, thus acting as an indirect driver of
greenwashing (\textit{ibid.}).

\subsection{Consequences of greenwashing}

The consequences of greenwashing can be dreadful as it can deceive consumers into believing they are making environmentally responsible choices when they are not. The resulting market confusion might lead to adverse choices from consumers. This undermines not ony the ability of consumers to drive positive environmental change through informed purchasing decisions but also,  mechanically, the undermining genuine environmental efforts of companies. Indeed, firms engaged in genuine sustainability efforts may be overshadowed by those that simply market themselves as green. This can lead to cynicism and scepticism among consumers about all environmental claims, making it more difficult for truly sustainable initiatives to gain traction. By diverting attention and resources towards ineffective or misleading solutions, greenwashing can slow down the progress towards genuine sustainability in various sectors.\\

\noindent Furthermore if greenwashing leads to the widespread adoption of products or practices that are not genuinely sustainable, it can result in continued or increased environmental damage. Consumers and investors may lose trust in companies caught greenwashing, potentially leading to a loss of market share, legal penalties, and a decrease in shareholder value. Increasing awareness of greenwashing can lead to stricter regulations and standards for environmental claims, which can have broader implications for industry practices and compliance costs. Greenwashing can reflect broader ethical issues within a company, indicating a willingness to deceive for profit. This can erode public trust in businesses and institutions more generally. These consequences highlight the importance of transparency, accountability, and accurate information in environmental marketing and corporate sustainability efforts.

\section{Crosswashing concept discussion}
\subsection{Crosswashing concept}

\begin{definition}
\textit{Crosswashing} refers to a corporate strategy employed by companies to bolster their sustainability image without making substantial alterations to their core business practices. This strategy is evident in the enhancement of their sustainability notation, particularly through investments in sustainable activities that are not associated with their primary operations when the core business is not inherently sustainable. Crosswashing involves the strategic cultivation of an impression of moral responsibility, providing protection against significant social risks, all while ensuring compliance with existing regulations.
\end{definition}

\noindent When sustainable initiatives undertaken by crosswashing firms significantly contribute to improving their ESG rating, these efforts should be labelled as crosswashing. However, if the initiatives do not have a substantial impact on the ESG rating, they do not fall under the category of crosswashing.\\

\noindent While greenwashing presents the practices of misleading consumers or the public with unsubstantiated environmental claims (see Section \ref{Greenwashing_overview}), Crosswashing, however, represents a new dimension of deceptive practices that goes beyond conventional greenwashing methods. Crosswashing manifests in diverse forms, all directed towards enhancing a company's sustainability rating while maintaining its core activities. In a series of studies, \cite{5d19c96f-caef-30d4-a4dc-9f338a61fce2} demonstrated how large corporations often ceremonially adopt but do not implement normatively mandated business practices, such as long-term incentive
programs (\cite{5d19c96f-caef-30d4-a4dc-9f338a61fce2} and \cite{69eec9d5-057f-311f-9038-36837815956d}) or stock repurchase programs (\cite{westphal2001decoupling}), to appear legitimate but still protect their core business
functions (\cite{Legitimacy}).\\

\noindent One common approach is when companies invest in sustainable sectors or products that have low costs and minimal direct relation to their principal business. For instance, oil companies may invest in social pillars of ESG frameworks, such as community development projects, to boost their overall ESG rating. While these investments may be commendable, they often overshadow the negative environmental impact of the company's core activities, such as oil extraction and fossil fuel production.
Another manifestation of crosswashing occurs when companies invest in costly sustainable sectors or products that have limited connection to their principal activities, but the invested amount is minimal compared to their core operations, creating an illusion of sustainability by selectively investing in renewable energies, for example, while continuing their significant oil production or other unsustainable harmful practices. By disproportionately highlighting their minimal investments, these companies seek to overshadow the broader environmental impact of their primary operations.\\

\noindent The lack of standardised methodologies and the divergence among existing sustainability rating systems (\cite{dimson2020divergent}) also provide fertile ground for crosswashing. Some companies exploit this situation by strategically abusing the non-standardisation and inconsistencies in methodologies to improve their sustainability ratings. By selecting frameworks or methodologies that align favourably with their existing practices, companies can manipulate their scores and present a distorted image of their sustainability efforts. One distinguishing aspect of crosswashing, particularly win comparison to other forms of greenwashing, is that companies engaging in crosswashing often comply with existing regulations and standards, minimising the risk of penalties. This compliance, however, should not be mistaken as a genuine commitment to sustainability. We express our deep concern at the rise of crosswashing, recognising that it has the potential to undermine genuine sustainability progress. Crosswashing allows companies to maintain the status quo while appearing environmentally responsible. It erodes trust in sustainability efforts and hampers the urgent need for transformative change.\\

\noindent However, it is crucial to distinguish between actual crosswashing and the risk of unfounded accusations of crosswashing. Accusations can arise due to factors such as inadequate data and data quality, methodologies that are difficult to comprehend, and overly broad definitions of greenwashing. It is essential to ensure that allegations are supported by robust evidence and thorough analysis, promoting a fair and accurate assessment of companies' sustainability practices.\\

\subsection{Crosswashing taxonomy}
\label{typology}
In this section, we explore the connection between crosswashing and greenwashing, drawing on the literature on greenwashing presented in Section \ref{Greenwashing_overview}.\\

In relation to typologies discussed in Section \ref{Greenwashingtypologyinliterature}, crosswashing is associated with \textit{decoupling behavior}. It diverges from a \textit{selective disclosure} approach as it enables a company to convey ostensibly positive information (a positive signal) ostensibly without being selective in its disclosures. Therefore, companies practising crosswashing often demonstrate a proclivity for voluntarily disclosing information. This adaptive strategy (\textit{decoupling behaviour}) facilitates a nuanced balance, allowing entities to maintain alignment with diverse environmental demands. As a result, the deliberate use of decoupling behaviour emerges as a pragmatic response to the complex challenges posed by the intersection of the influences of financial drivers and stakeholders influences.\\

\noindent Through the adoption of \textit{decoupling behaviour},  crosswashing contributes to the enhancement of the legitimacy of firms, particularly in terms of cognitive legitimacy, moral legitimacy, and pragmatic legitimacy. \quotestart The research provides a strong empirical evidence of the use of decoupling as a critical strategy that enables firms to maintain a degree of 'fit' with both technical–economic pressures from their economic environment and cognitive,
normative, and regulatory pressures from their institutional environment" (\cite{Legitimacy}). Crosswashing elevates \textit{cognitive legitimacy} by establishing the organisational characteristic or practice as normatively acceptable within its environment to the extent that it is perceived as \quotestart natural” (see \cite{Hannan1986} and \cite{Legitimacy}). This phenomenon aligns with the definition provided by \cite{Legitimacy}, wherein \quotestart cognitive legitimacy is an extension of sociopolitical legitimacy that occurs when there is such a high degree of congruence or acceptance between the normative expectations of the organisation and its environment that they are unquestioned or taken for granted."\\

\noindent Regarding \textit{moral legitimacy}, crosswashing augments this aspect as firms making green and social claims, through the crosswashing strategy, will present themselves as aligned with its practical impact. This alignment is consistent with the definition in \cite{Legitimacy}, citing \cite{scott1995symbols}, which characterizes "normative legitimacy" as a measure of congruence or harmony between the actions, characteristics, and structure of the organisation and the beliefs and cultural values of the broader social environment in which it operates.\\

\noindent \textit{Pragmatic legitimacy} emerges as "the result of self-interested calculations of the organisation’s key stakeholders, and it is based on stakeholders' perceptions of their personal benefit deriving from corporate activities and communication" (\cite{https://doi.org/10.1002/bse.1912}). We argue that \textit{pragmatic legitimacy} may be the central element among the three legitimacy strategies in the context of \textit{Crosswashing}. Indeed, financial and economic outcomes constitute the primary drivers behind both greenwashing and crosswashing. Crosswashing, in particular, offers firms a valuable opportunity to derive benefits from controversial activities while minimising the risk of being perceived as irresponsible. Thus, we acknowledge \textit{pragmatic legitimacy} as a pivotal strategy for firms, especially in the context of crosswashing. However, this acknowledgement does not align with the observation of \cite{Legitimacy}, where it is noted that \quotestart this specific category of legitimacy is perhaps the least common to other typologies and, in fact, has been rejected by ecologists who argue that the test is not one of legitimacy at all but rather describes a degree of organisational learning (\cite{AldrichRuef2006})".

\subsection{The common \textit{characteristics} of crosswashing and greenwashing}
\label{characteristics}

In this section, we will assess the shared traits between greenwashing and crosswashing, drawing from the details provided in Section \ref{Characteristics_of_Green_washing}.\\

\noindent \textbf{\textit{Claim greenwashing:}}  Regarding the concept of \quotestart claim greenwashing", the initial category, referred to as \quotestart deceptive claims", described as \quotestart lying, lying by omission, or lying through lack of clarity" (\cite{de2020concepts}), does not align with the characteristics of \textit{crosswashing}. This is because it concentrates on the accuracy of shared information, such as truth data. In contrast, the \textit{crosswashing} strategy involves the construction of favourable sustainable data, as explained earlier.\\

\noindent The second category of \quotestart claim greenwashing," known as the "claim type," comprises four types (a, b, c, d)\footnote{Types (a), (b), (c), (d). Type (e) represents a combination of these four types.}, as detailed in 
Section \ref{Characteristics_of_Green_washing}. Within the framework of these four types, it is essential to note that (b) and (c) distinctly embody the traits associated with crosswashing.\\

\noindent In fact, claim (a), which pertains to "product orientation claims", does not encompass the characteristics of crosswashing as it (claim (a)) challenges the accuracy of a firm's assertions regarding the environmental friendliness of its products or services (\cite{doi:10.1525/cmr.2011.54.1.64}). crosswashing, on the other hand, does not face this risk since crosswashing firms provide accurate information about their products and activities. Claim (d), referred to as \quotestart environmental fact", does not aaply to crosswashing because, within the context of crosswashing, all independent statements made by firms are genuinely factual.\\

\noindent The two remaining types of claims collectively characterise crosswashing. Specifically, the claim associated with the \textit{orientation process} (claim (b)) stands out as a distinctive feature of crosswashing. That is, in the context of greenwashing, firms aim to perpetuate their non-sustainable activities within their overall processes, without ensuring that the entirety of the firm's processes will align with sustainable targets. 
Claim (c), referred to as \quotestart image orientation claims", also characterises crosswashing. This is because in the realm of crosswashing, companies engaging in greenwashing may assert that they are advocating for a specific aspect of sustainable objectives; however, the true objective is to divert attention from their primary non-sustainable activities.\\

 \noindent \textit{\textbf{The thirteen sins of greenwashing}}: The second classification, known as the \textit{thirteen sins of greenwashing}, includes \textit{six} that characterise \textit{crosswashing}, specifically (6) lesser of two evils, (8) false hopes, (10) broken promises, (11) injustice, (12) hazardous consequences, and (13) profits over people and the environment. Sin (9), the sin of fearmongering, could potentially be linked to \textit{crosswashing} as it seeks to persuade people, stakeholders, and authorities that the continuation of undesirable activities is essential for economic activity, for example.\\

\noindent Sins (6), (8), (10), and (12) specifically point directly  to \textit{crosswashing}. Fuel-efficient sport-utility vehicles exemplify the sin of the lesser of two evils, as their claim of being environmentally friendly within the SUV category may distract consumers from the broader environmental impacts associated with fuel consumption. False hope involves creating false hope that the firm's strategy will benefit the environment, even when such benefits are uncertain. The broken promise aspect emerges from sustainable activities used by firms for crosswashing, promising an effective way to enhance certain environmental or social aspects. For instance, some companies finance social projects in rural areas of poor countries to offset their negative activities, yet these projects often yield only short-term and negligible effects. This latter example also represents a form of injustice committed by firms engaging in crosswashing, as they frequently target vulnerable populations.\\ 

\noindent The following sins do not align with crosswashing because they pertain to distinct types of misleading or deceptive practices that are not central to the concept of crosswashing:  (1) the sin of the hidden trade-off, (2) the sin of no proof, (3) the sin of vagueness, (4) the sin of worshipping false labels, (5) the sin of irrelevance, and (7) the \textit{sin of fibbing} do not align with crosswashing because they pertain to distinct types of misleading or deceptive practices that are not central to the concept of crosswashing. Notably, the \textit{sin of the hidden trade-off}, defined as \quotestart a claim suggesting that a product is ‘green’ based on a narrow set of attributes without attention to other important environmental issues" (\cite{de2020concepts}), does not apply to crosswashing. This is because the characteristics of green and social products and services emphasised by firms engaging in crosswashing are often real, even if the overall environmental impact of the firm's activities is less sustainable than implied. The sin of no proof, which is characterised as \quotestart an environmental claim that cannot be substantiated by easily accessible supporting information or by a reliable third-party certification" (\cite{de2020concepts}), does not accurately depict crosswashing either. Crosswashing is expected to provide credible data as it is one of its components, but this doesn't necessarily imply the complete supporting information or third-party certification in its entirety. The \textit{sin of vagueness}, described as making an unclear or imprecise environmental claim without a clear meaning (\cite{de2020concepts}), also does not aptly characterise crosswashing. Crosswashing is constructed to present its claims with a degree of credibility, yet this does not guarantee complete clarity or precision; instead, the information provided may be unintentionally ambiguous or vague.
The \textit{sin of worshipping false labels}, characterised as \quotestart a product that, through a false suggestion or certification-like image, misleads consumers into thinking that it has been through a legitimate green certification process, an example being a paper towel with packaging featuring a certification-like image claiming that the product \textit{fights global warming}" (\cite{de2020concepts}), is not an exact representation of crosswashing. Crosswashing is structured to present its claims with a degree of credibility and does not involve creating a false suggestion or using certification-like images to mislead consumers into thinking that the product has undergone a legitimate green certification process.
The \textit{sin of fibbing}, defined as \quotestart environmental claims that are simply false' (\cite{de2020concepts}),  diverges from crosswashing due to the same underlying causes as the sins mentioned above.\\

 \textbf{\textit{Firm-greenwashing transgressions:}} The third classification, \textit{firm-greenwashing transgressions}, consists of the \textit{five} distincts manifestations of \textit{greenwashing transgressions} described in Section \ref{Characteristics_of_Green_washing}. These manifestations include (i) dirty business, (ii) ad bluster, (iii) political spin, (iv) It is the law, stupid!, and (v) Fuzzy reporting. 
 Manfestations (ii), (iv), and (v) are not indicative of \textit{crosswashing}, as crosswashing firms are characterised by providing accurate information and compliance with regulations, as cited above. Only (ii) exhibits a tenuous connection with crosswashing, as it involves partially the presentation of alternative programs unrelated to the main sustainability concern.  Regarding (iii), political spin, we assert that it does not align with the concept of \textit{crosswashing} in this particular paper. In reality, political pressures are substantial and play a significant role, which large \textit{crosswashing} firms can leverage without facing disapproval from stakeholders.\\

\noindent The term \quotestart dirty business" (i),  as articulated by \cite{urban2019bank}, refers to corporations falling short of fundamental standards in sustainable corporate practices across ESG dimensions. This definition aligns with the conceptualisation provided by \cite{contreras2017fuzzy}, as discussed in Section \ref{Characteristics_of_Green_washing}. Diverse perspectives on the concept of "dirty business" are explored in various literature, including in the works of \cite{punch1996dirty}, \cite{mitchell2002dirty}, and \cite{lambrechts2016environmental}. Dirty business encompasses multiple dimensions. Firstly, it involves corporate misconduct concerning environmental damage, severe corruption, and human rights violations. Secondly, it relates to the types of goods produced by companies, encompassing items like tobacco, coal, nuclear weapons, and cluster munitions. Thirdly, it extends to the serious violation of fundamental ethical norms, exemplified by scenarios where resource extraction and exploration in certain African regions fail to bring benefits to the local population (\cite{urban2019bank}). Hence, the term \textit{dirty business} encompasses a range of activities, including both greenwashing and crosswashing practices.

\subsection{Crosswashing drivers}
\label{Drivers}
\noindent Based on the content in Section \ref{GreenwashingDrivers}, we will explore whether the drivers of crosswashing are analogous to those of greenwashing.\\

\noindent \textbf{Individual-level drivers:} We acknowledge that individual-level greenwashing actors do not necessarily translate into drivers of crosswashing. Firstly, the \quotestart narrow decision framing" tendency is incongruent with the dynamics of crosswashing, which involves interdisciplinary exchange to define effective strategies that are not easily detectable by stakeholders. Crosswashing demands a comprehensive approach that goes beyond compartmentalised decision-making, requiring collaboration across various domains to create strategies that are strategically sound yet apparent to stakeholders.\\

\noindent Secondly, the \quotestart hyperbolic intertemporal discounting" direct applicability to crosswashing is not evident. In the context of crosswashing, individuals need to demonstrate patience and a willingness to invest in long-term strategies that may not yield immediate results. Crosswashing involves a more strategic and enduring perspective where individuals are required to navigate complex interdisciplinary challenges and invest in strategies that may take time to unfold.\\

\noindent Lastly, \quotestart optimistic bias", in connection to crosswashing,  is not straightforward since in the context of crosswashing, individuals often adopt a more realistic approach. They recognise the complexities involved, assess risks accurately, and assign the appropriate weight to each decision to ensure the success of crosswashing strategies. The realism inherent in crosswashing strategies sets it apart from the optimistic bias associated with greenwashing, reflecting a more nuanced and strategic decision-making process.\\

\noindent Hence, the individual-level drivers identified, such as narrow decision framing, hyperbolic intertemporal discounting, and optimistic bias, primarily pertaining to greenwashing behaviours, may not inherently serve as direct drivers for crosswashing, given the unique considerations and circumstances associated with each phenomenon.\\

\noindent \textbf{Organisational level of drivers:} 
Crosswashing provides a strategic avenue for companies that may face scrutiny from environmentalists and NGOs for potential greenwashing. By engaging in crosswashing activities, these companies can enhance their overall ESG (Environmental, Social, and Governance) score over time. This improvement in ESG metrics serves as a means to safeguard or justify their primary non-sustainable activities, offering a proactive approach to address concerns raised by environmental advocates and non-governmental organisations. However, the specific dynamics of crosswashing, specifically those involving interdisciplinary strategies, may introduce nuances in how these characteristics impact the likelihood of being targeted.
Concerning the \noindent \textit{organisational inertia}, which involves a gap between environmentally friendly intentions discussed by management and actual implementation, is a driver for greenwashing. In the context of crosswashing, where a company engages in interdisciplinary strategies, the inertia might manifest differently. Crosswashing may necessitate overcoming inertia through effective coordination and collaboration across various organisational domains. In the context of the \textit{effectiveness of intra-firm communication} drivers, companies that grapple with ineffective communication between departments and sections face an increased likelihood of resorting to greenwashing practices. Conversely, the triumph of crosswashing strategies relies on the organisation's ability to communicate and coordinate seamlessly across diverse departments and disciplines.\\

\noindent \textbf{Market external drivers:} 
In the context of crosswashing, market external drivers consumer demand, investor demand, and competitive pressure play a strategic role in helping companies manage external expectations and diminish pressure. Rather than feeling compelled to embellish their images, companies engaged in crosswashing strategically leverage these drivers to enhance their sustainability image without making substantial changes to their core business practices. Crosswashing becomes a proactive approach for companies to align with consumer and investor preferences and stay competitive in the market.\\

\noindent \textbf{Regulatory context: }  In the context of crosswashing, companies strategically leverage the regulatory environment as a critical driver to enhance their sustainability image without facing significant punitive consequences. Unlike greenwashing, where limited regulatory penalties drive deceptive practices, crosswashing involves a proactive approach. Companies operating under crosswashing use the constraints of regulations to their advantage, ensuring compliance while strategically managing external expectations. This approach transforms the regulatory context into a tool for aligning with regulations and simultaneously enhancing public perception, offering companies a nuanced strategy to navigate the intricate landscape of sustainability and regulatory compliance.

\section{Use case: Applying a multi-criteria decision-making (MCDM) approach to analyse the TotalEnergy company taxonomy data}
\label{usecase}

In this case study, we adhere to the  MCDM methodology described in \cite{roszkowska2013rank}. MCDM is an approach designed to assist decision-makers in evaluating, prioritising, and selecting from various conflicting alternatives and criteria. MCDM is also known by various terms such as multiple-criteria decision analysis (MCDA), multi-criteria decision aiding (MCDA), multi-attribute decision analysis (MADA), multiple objective decision analysis (MODA), and Single Participant-Multiple Criteria Decision Making (SPMC) (\cite{RePEc:gam:jsusta:v:9:y:2017:i:2:p:287-:d:90524})\footnote{See (\cite{koksalan2011multiple}, \cite{ishizaka2012ahp}) (\cite{GORSEVSKI2012287}) for more details}.\\

\noindent \quotestart Multicriteria decision making (MCDM) refers to screening, prioritising, ranking or selecting the alternatives based on human judgment from among a finite set of ` alternatives in terms of the multiple 
usually conflicting criteria" (\cite{roszkowska2013rank}). 
We leverage Multicriteria Decision Making (MCDM) in line with our primary objective of systematically ranking alternatives based on specific criteria to derive a comprehensive score. This score is instrumental in establishing a connection between the alternatives and the main activities of the company. It is noteworthy that these alternatives correspond to the activities of the company, although they may not represent its core or main activity. This nuanced evaluation allows us to gauge the alignment of alternatives with the company's diverse operations and strategic focus, aiding in the detection and quantification of crosswashing levels. The general steps involved in MCDA are as follows:

\noindent we have $I$ alternatives (\(A_i, i \in [0, I]\), ) and $J$ criteria (\(C_i, i \in [0, I]\), ). 

\begin{equation}
    S_i = \sum_{j} w_j \cdot V_{ij}
\end{equation}

\noindent where, \(S_i\) represents the overall score for alternative \(A_i\). It is the sum of the product of each criterion's weight (\(w_j\)) and the normalised value of the alternative for that criterion (\(V_{ij}\)) where \(\sum_{j} w_j = 1\).\\

In our case study, the alternatives represent activities as outlined in Table \ref{tab11}. The criteria used for evaluation are presented in Table \ref{tab6} along with their respective weights. It's important to note that these weights are arbitrarily assigned by the authors. The objective here is to showcase the methodologies rather than provide definitive scores.

\subsection{Data}

We use data from \cite{totalenergies_ESG_Databook2022}\footnote{The utilization of data from TotalEnergy in this context is purely for illustrative purposes and does not constitute a critique of the company's practices. We acknowledge and appreciate TotalEnergy's commitment to transparency, and our use of their data is intended solely to provide a tangible example in the discussion at hand.}, which are available in both Excel and PDF formats, with the former as our preferred. . Within the Excel file, there are multiple sheets, and we specifically utilise the \quotestart Taxo EN" sheet. We utilise a specific section of the \quotestart TURNOVER RATIOS - Controlled scope" table, specifically corresponding to \quotestart A.1. Environmentally sustainable activities (taxonomy-aligned)" and a similar section from the \quotestart CAPEX RATIOS - Controlled scope" table.\\

\noindent In the initial table (Table \ref{tab10}), the total turnover value of taxonomy-eligible activities amounts to \SI{19865} M\$ US, constituting \SI{7.5}{\percent} of the overall activities. Of this, \SI{3466} M\$ US, representing aligned activities, are considered in this study. In the second table (Table \ref{tab9}), the total capex value of taxonomy-eligible activities amounts to \SI{3168} M\$ US, constituting \SI{17.35}{\percent} of the overall activities. Of this, \SI{2652}  M\$ US, representing aligned activities, are considered in this study.\\

\noindent Table \ref{tab11} provides the \textit{Link} and \textit{Contribution} levels for each of the selected activities of TotalEnergy activities. Each value can range from 0 to 5. A value of 0 for \textit{Link} indicates that the activity has no connection with the main TotalEnergy activity. Regarding \textit{Contribution}, a value of 0 signifies that the activity does not contribute to reducing the main activity (oil and gas extraction). Conversely, a \textit{Link} value of 5 implies a close association with the main activity, while a \textit{Contribution} value of 5 signifies a substantial contribution to transitioning to a low-carbon activity, reducing the primary activity. The values were chosen by the authors with justification in the same table (Table \ref{tab11}) and require evaluation in collaboration with industry professionals. The objective is to present a potential approach rather than providing a final assessment of TotalEnergy. Table \ref{tab1} provides a description of the variables. Tables \ref{tab9}
and \ref{tab10} present selected columns from the two data tables used, notably \quotestart CAPEX RATIOS - Controlled scope" and \quotestart TURNOVER RATIOS - Controlled scope". Tables \ref{tab2}  and \ref{tab3} gives data from tables \ref{tab9} and \ref{tab10} enriched by table \ref{tab11}, grouped by \textit{Link} and \textit{Contribution}, respectively.\\

\noindent Tables \ref{tab4} and \ref{tab5} merge \footnote{After the merge, we still have only 14 activities. Table \ref{tab9} contains 19 activities, and Table \ref{tab9} contains 17.} the data from  Tables \ref{tab9}  and \ref{tab10}, enriched by Table \ref{tab11} and grouped by \textit{Link} and \textit{Contribution}, respectively. The data from these merged tables is used in the MCDM approach.\\

\noindent The choice of using \textit{CapEx (Capital Expenditure)} and \textit{Turnover} variables in our study is based on their significance in assessing the financial and operational aspects of activities within TotalEnergy. These variables provide valuable insights into the investment and revenue aspects of each activity, offering a comprehensive view of their financial performance. \textit{CapEx} helps evaluate the financial commitment and investment made by TotalEnergy in each activity. It reflects the resources allocated for the acquisition, upgrade, or maintenance of assets, indicating the company's strategic investment decisions. \textit{Turnover}, or revenue generated by each activity, is a crucial financial metric. It helps assess the economic viability and contribution of each activity to TotalEnergy's overall financial performance. Activities with higher turnover may indicate greater financial sustainability and contribution to the company's revenue stream.\\

\subsection{Results and discussion of MCDM approaches}

In this section, we apply the MCDM approach twice. Firstly, considering \textit{Link, Contribution, CapEx (Capital Expenditure) in Millions of Dollars, Turnover in Millions of Dollars} as criteria, taking into account CapEx and Turnover as proxies for the activities' investment costs and revenues. Secondly, we use only \textit{Link} and \textit{Contribution}, neglecting the financial aspect.

\subsubsection{MCDM with Criteria: \textit{Link, Contribution, CapEx M\$, Turnover M\$}}
\label{mcdmcpaex}

\noindent Table \ref{tab7} illustrates the\textit{Normalised Weighted sum} for each alternative (activity). The three most representative activities aligning with Totalenergy's main objectives are \textit{\quotestart Electricity generation from wind power"} at $100\%$, \textit{\quotestart Electricity generation using solar photovoltaic technology"} at $91\%$, and \textit{\quotestart Installation, maintenance and repair of renewable energy tech"} at $69\%$. On the contrary, \textit{\quotestart Professional services related to energy performance of buildings"} and \textit{\quotestart Anaerobic digestion of bio-waste"} only exhibit $31\%$ and $30\%$ alignment with Totalenergy's main activity. Importantly, \textit{\quotestart Manufacture of plastics in the primary form"} is not aligned with Totalenergy's main activity, presenting a $0\%$ alignment.\\

\noindent The \textit{Average Normalised Weighted Sum} is 53.20\%. That means that TotalEnergy's environmental note currently calculated on taxonomy aligned activity should be deteriorated by 46.8\% (100-53.20\%).\\

\subsubsection{MCDM with Criteria: \textit{Link, Contribution}}
\label{mcdmcpaex2}
Table \ref{tab8} presents the outcomes of the analysis, revealing a substantial shift when considering only the Link and Contribution criteria for activities. Notably, both \textit{\quotestart Manufacture of energy efficiency equipment for buildings"} and \textit{\quotestart Electricity generation from wind power"} exhibit a complete alignment with Totalenergy's activity, scoring $100\%$. Following closely are \textit{\quotestart Manufacture of low carbon technologies for transport"}, \textit{\quotestart Electricity generation using solar photovoltaic technology"}, and \textit{\quotestart Storage of electricity"}, registering alignment percentages of $87\%$, $80\%$, and $80\%$, respectively. On the other end of the spectrum, \textit{\quotestart Professional services related to energy performance of buildings"} and \textit{\quotestart Anaerobic digestion of bio-waste"} maintain their distance from Totalenergy's activity, as discussed in Section \ref{mcdmcpaex}, recording alignment percentages of $40\%$ and $47\%$, respectively.  \textit{\quotestart Manufacture of plastics in the primary form"} still demonstrates no alignment with Totalenergy's primary activity, scoring $0\%$, as  mentioned  previously in Section \ref{mcdmcpaex}.\\

\noindent The \textit{Average Normalised Weighted Sum} is  66.20\%, which  means that TotalEnergy's environmental note currently calculated on taxonomy aligned activity should have deteriorated by 33.8\%. 

\subsubsection{Interpretation of results}

The deterioration level in Section \ref{mcdmcpaex2} (33.8\%) is lower than that in Section \ref{mcdmcpaex} (53.2\%). This suggests that financial drivers may not be the primary influence behind Totalenergy's strategies for sustainable activities. Additionally, this observation is partially supported by the results in Tables \ref{tab2} and \ref{tab3}. Table \ref{tab2} indicates that over 90\% of CapEx is concentrated in activities with Link levels 3 and 4 (\textit{CapEx Perc2}). In Table \ref{tab3}, 84\% (at level 3) of Totalenergy's activities contribute to the reduction of its oil and gas extraction.\\

\noindent Activities contributing to the reduction of oil and gas extraction are more profitable, as evidenced by the high percentage (98\%) in Table \ref{tab3}, column \textit{Turnover Perc2}. On the other hand, activities linked to Totalenergy's main activity are less profitable, accounting for only 13\% of activities according to Table \ref{tab2}, column \textit{Turnover Perc2}.

\section{Concluding remarks}

Addressing \textit{crosswashing} requires a multi-faceted approach involving regulatory action, industry cooperation, and consumer awareness. Stricter regulations should be implemented to guarantee the effectiveness of companies' sustainable development actions. Regulators need to scrutinise companies' investments, demand clarity on the relevance of these investments to core activities, and discourage deceptive practices that overshadow environmental impacts.\\

\noindent This paper endeavours to make a meaningful contribution to the literature in this regard by putting forth a comprehensive evaluation of \textit{crosswashing}. 
As mentioned earlier, \textit{crosswashing} is a subset of greenwashing, and by examining their shared attributes in Table \ref{cwch}, we can infer that the characteristics  of \textit{crosswashing} (discussed in Section \ref{characteristics}) align with those labelled as \textit{strategic}. Consequently, these characteristics suggest a focus on long-term considerations.\\

\noindent The \textit{strategic} attribute of \textit{crosswashing} elucidates its alignment with the typology of \textit{decoupling behaviour} expounded in Section \ref{typology}, primarily through its reinforcement of the three types of \textit{legitimacy}, which is discussed in the same section. This alignment stems from the fact that crosswashing enhances \textit{cognitive legitimacy} by establishing organisational characteristics or practices as normatively acceptable within its environment, to the extent that they are perceived as "natural." Additionally, \textit{moral legitimacy} contributes to the alignment by augmenting the congruence or harmony between the actions, characteristics, and structure of the organisation and the beliefs and cultural values of the broader social environment in which it operates.
Furthermore, \textit{pragmatic legitimacy} is heightened as crosswashing enhances financial and economic outcomes, which constitute the primary drivers behind firms' objectives.\\

\noindent Regarding \textit{crosswashing drivers} (Section \ref{Drivers}), it is crucial to underscore that in the \textit{strategic} context, \textit{crosswashing} does not align with the \textit{individual drivers} of greenwashing due to their short-term attributes and non-strategic characteristics. Instead, \textit{crosswashing} corresponds more closely to \textit{organisational-level drivers}, \textit{market external drivers}, and the \textit{regulatory context}, which necessitate strategic policies.\\

\noindent The approach proposed in Section \ref{usecase} offers an initial framework to gauge the impact of \textit{crosswashing} on \textit{firms' ESG notation}. By scrutinising green firms' activities through variables such as their connection to the main business, contribution to mitigating the main activity's impact, investment costs, and revenue, it lays the foundation for a comprehensive evaluation. However, to capture the full spectrum of \textit{crosswashing} effects, additional variables, particularly those capturing the dynamic nature of these activities and their influence on core business functions, must be considered. Although this case study concentrates solely on the \textit{environmental} pillar of ESG due to data limitations, its methodology provides a robust and holistic approach to understanding the intricate relationship between crosswashing strategies and the  ESG performance of firms.\\

\noindent The findings derived from the case study indicate a notable overestimation in current ESG notations. This overestimation, however, is contingent upon the specific industry sectors and the size of the companies involved. It underscores the necessity for a nuanced understanding that recognises variations in the impact of crosswashing across different sectors and scales of enterprises. Consequently, there is a compelling argument for a revision of ESG notation methodologies to incorporate and account for the nuanced effects of crosswashing. Such revisions should acknowledge sector-specific dynamics and the influence of company size, fostering a more accurate and insightful assessment of ESG performance that reflects the diverse strategies and impacts observed in the context of sustainability practices.\\
 
\noindent In conclusion, crosswashing represents a concerning development in the realm of sustainability. This manipulative strategy allows companies to enhance their sustainability ratings without addressing the core environmental impact of their operations. Urgent action is needed to expose and deter crosswashing, ensuring that sustainability efforts are genuine, transparent, and transformative. By doing so, we can safeguard the integrity of sustainability initiatives and foster a healthier, more sustainable future for all.

\begin{xltabular}{\textwidth}{@{}clX@{}}
    \caption{Common Characteristics of Crosswashing and Greenwashing } \label{cwch} \\ 
    \toprule
    Characteristics & Sub-characteristics & y/n \\
    \midrule
    \endhead
    \bottomrule
    \endfoot
    Deceptive Claim & Deceptive Claim & n \\
    
    Claim Type & Product Orientation Claims & n \\
     & Orientation Process & y \\
     & Image Orientation Claims & y \\
     & Environmental Fact & n \\
    
    The Thirteen Sins of Greenwashing & The Sin of the Hidden Trade-off & n \\
    The Thirteen Sins of Greenwashing & Sin of No Proof & n \\
     & Sin of Vagueness & n \\
     & Worshiping False Labels & n \\
     & The Sin of Irrelevance & n \\
     & Sin of Lesser of Two Evils & y \\
     & Sin of Fibbing & n \\
     & False Hopes & y \\
     & Broken Promises & y \\
     & Injustice & y \\
     & Hazardous Consequences & y \\
     & Profits over People and the Environment & y \\
    
    Firm-Greenwashing Transgressions & Dirty Business & y \\
     & Ad Bluster & n \\
     & Political Spin & n \\
     & It is the Law, Stupid! & n \\
     & Fuzzy Reporting & n \\
\end{xltabular}

\begin{xltabular}{\textwidth}{@{}l X@{}}

    \caption{Description of Variables} \label{tab1} \\
    \toprule
    \textbf{Variable} & \textbf{Description} \\
     \midrule
    \endhead
    \bottomrule
    \endfoot
     Link & Level of connection between activity and main activity of TotalEnergies’. \\
    Contribution & Level of contribution of an activity to reduce TotalEnergy's main activity. \\
    CapEx M\$ & Capital expenditure in millions of dollars incurred by the activity. \\
    CapEx Perc & Percentage of capital expenditure (CapEx) in relation to the total. \\
    CapEx Perc2 & Percentage of capital expenditure (CapEx) within selected activities. \\
    Turnover M\$ & Turnover in millions of dollars generated by the activity. \\
    Turnover Perc & Percentage of turnover in relation to the total. \\
    Turnover Perc2 & Percentage of Turnover within selected activities. \\
\end{xltabular}

\begin{xltabular}{\textwidth}{@{}ccccccc@{}}
    \caption{Capex and Turnover Grouped by Link Level} \label{tab2} \\
    \toprule
    Link & CapEx M\$ & CapEx\_Perc & CapEx Perc2 & Turnover M\$ & Turnover Perc & Turnover Perc2 \\
    
    \midrule
    \endhead
    \bottomrule
    \endfoot
    
    0 & 4 & 0.000219 & 0.150830  & 352 & 0.001337 & 10.155799 \\
    1 & 80 & 0.004382 & 3.016591  & 1503 & 0.005708 & 43.364108 \\
    2 & 150 & 0.008216 & 5.656109  & 1165 & 0.004424 & 33.612233 \\
    3 & 1446 & 0.079202 & 54.524887  & 446 & 0.001694 & 12.867859 \\
    4 & 972 & 0.053240 & 36.651584 & & &  \\ 
    
\end{xltabular}

\begin{xltabular}{\textwidth}{ccccccc}
    \caption{Capex and Turnover Grouped by Contribution} \label{tab3} \\
    \toprule
    Contibution & CapEx M\$ & CapEx Perc & CapEx Perc2 & Turnover M\$ & Turnover Perc & Turnover Perc2 \\
    \midrule
    \endhead
    0 & 21 & 0.001150 & 0.791855 & 47 & 0.000178 & 1.356030 \\
    1 & 114 & 0.006244 & 4.298643 & & & \\
    2 & 292 & 0.015994 & 11.010558 & 1578 & 0.005993 & 45.527986 \\
    3 & 2225 & 0.121871 & 83.898944 & 1841 & 0.006992 & 53.115984 \\
    \bottomrule
\end{xltabular}

\begin{xltabular}{\textwidth}{ccccccc}
    \caption{Capex and Turnover After Merge Grouped by Link Level} \label{tab4} \\
    \toprule
    Link & CapEx M\$ & CapEx Perc & Turnover M\$ & Turnover Perc & CapEx Perc2  & Turnover Perc2 \\
    \midrule
    \endhead
    1 & 30 & 0.001643 & 352 & 0.001337 & 1.182033 & 10.404966 \\
    2 & 150 & 0.008216 & 1430 & 0.005431 & 5.910165 & 42.270174 \\
    3 & 1406 & 0.077012 & 1155 & 0.004386 & 55.397951 & 34.141295 \\
    4 & 952 & 0.052144 & 446 & 0.001694 & 37.509850 & 13.183565 \\
    \bottomrule
\end{xltabular}

\begin{xltabular}{\textwidth}{ccccccc}
    \caption{Capex and Turnover After Merge Grouped by Contribution} \label{tab5} \\
    \toprule
    Contibution & CapEx M\$ & CapEx Perc & Turnover M\$ & Turnover Perc & CapEx Perc2  & Turnover Perc2 \\
    \midrule
    \endhead
    0 & 21 & 0.001150 & 47 & 0.000178 & 0.827423 & 1.389299 \\
    2 & 292 & 0.015994 & 1497 & 0.005685 & 11.505122 & 44.250665 \\
    3 & 2225 & 0.121871 & 1839 & 0.006984 & 87.667455 & 54.360035 \\
    \bottomrule
\end{xltabular}

\begin{xltabular}{\textwidth}{p{4.5cm}cc}
    \caption{Weights for Each Criterion} \label{tab6} \\
    \toprule
    Criterion & Weight \\
    \midrule
    CapEx & 0.3 \\
    Turnover & 0.2 \\
    Link level & 0.3 \\
    Contibution & 0.2 \\
    
    \bottomrule
\end{xltabular}

\begin{xltabular}{\textwidth}{p{6.5cm}cc}
    \caption{Multi-criteria Analysis Results - With CapEx and Turnover} \label{tab7} \\
    \toprule
    Activities & Weighted Sum & Normalised Weighted Sum (\%) \\
    \midrule
    \endhead
    Electricity generation from wind power & 0.77 & 100.00 \\
    Electricity generation using solar photovoltaic technology & 0.71 & 91.35 \\
    Installation, maintenance, and repair of renewable energy tech. & 0.54 & 69.08 \\
    Manufacture of energy efficiency equipment for buildings & 0.50 & 64.72 \\
    Manufacture of low carbon technologies for transport & 0.49 & 62.82 \\
    Manufacture of batteries & 0.46 & 59.24 \\
    Storage of electricity & 0.43 & 54.43 \\
    Infrast. enabling low-carbon road transport and public transport & 0.39 & 49.45 \\
    Manufacture of biogas/biofuels for use in transport & 0.37 & 47.59 \\
    Installation, maintenance and repair of charging stations for electric vehicles & 0.34 & 42.84 \\
    Production of heat/cool using waste heat & 0.33 & 42.26 \\
    Professional services related to energy performance of buildings & 0.25 & 30.74 \\
    Anaerobic digestion of bio-waste & 0.24 & 30.32 \\
    Manufacture of plastics in primary form & 0.01 & 0.00 \\
    \bottomrule
\end{xltabular}

\begin{xltabular}{\textwidth}{p{6.5cm}cc}
    \caption{Multi-criteria Analysis Results - Without CapEx and Turnover}
    \label{tab8} \\
    \toprule
    \textbf{Activities} & \textbf{Weighted Sum} & \textbf{Normalised Weighted Sum (\%)} \\
    \midrule
    \endhead
    \endfoot
    
    Manufacture of energy efficiency equipment for building & 0.50 & 100.00 \\
    Electricity generation from wind power & 0.50 & 100.00 \\
    Manufacture of low carbon technologies for transport & 0.43 & 86.67 \\
    Electricity generation using solar photovoltaic technology & 0.40 & 80.00 \\
    Storage of electricity & 0.40 & 80.00 \\
    Manufacture of batteries & 0.33 & 66.67 \\
    Manufacture of biogas/biofuels for use in transport & 0.33 & 66.67 \\
    Production of heat/cool using waste heat & 0.33 & 66.67 \\
    Infrast. enabling low-carbon road transport and public transport & 0.33 & 66.67 \\
    Installation, maintenance and repair of charging stations for electric vehicles & 0.33 & 66.67 \\
    Installation, maintenance and repair of renewable energy tech. & 0.30 & 60.00 \\
    Anaerobic digestion of bio-waste & 0.23 & 46.67 \\
    Professional services related to energy performance of building & 0.20 & 40.00 \\
    Manufacture of plastics in primary form & 0.00 & 0.00 \\
    \bottomrule
\end{xltabular}
\begin{xltabular}{\textwidth}{p{6cm}ccc}

    \caption{CAPEX RATIOS - Controlled Scope (First Four Columns)} \label{tab9}\\
    \toprule
    Activities & Code & CapEx M\$ & CapEx Perc \\
    \midrule
    \endhead
    \bottomrule
    \endfoot

    Afforestation & 1.1 & 4 & 0.0\% \\
    Manufacture of low carbon technologies for transport & 3.3 & 13 & 0.1\% \\
    Manufacture of batteries & 3.4 & 36 & 0.2\% \\
    Manufacture of energy efficiency equipment for buildings & 3.5 & 1 & 0.0\% \\
    Manufacture of organic basic chemicals & 3.14 & 37 & 0.2\% \\
    Manufacture of plastics in primary form & 3.17 & 21 & 0.1\% \\
    Electricity generation using solar photovoltaic technology & 4.1 & 1,060 & 5.8\% \\
    Electricity generation from wind power & 4.3 & 938 & 5.1\% \\
    Storage of electricity & 4.10 & 85 & 0.5\% \\
    Manufacture of biogas/biofuels for use in transport & 4.13 & 54 & 0.3\% \\
    Production of heat/cool using waste heat & 4.25 & 1 & 0.0\% \\
    Anaerobic digestion of bio-waste & 5.7 & 18 & 0.1\% \\
    Underground permanent geological storage of CO2 & 5.12 & 20 & 0.1\% \\
    Infrast. enabling low-carbon road transport and public transport & 6.15 & 167 & 0.9\% \\
    Construction of new buildings & 7.1 & 3 & 0.0\% \\
    Installation, maintenance and repair of charging stations for electric vehicles & 7.4 & 3 & 0.0\% \\
    Installation, maintenance and repair of renewable energy tech. & 7.6 & 132 & 0.7\% \\
    Acquisition and ownership of buildings & 7.7 & 50 & 0.3\% \\
    Professional services related to energy performance of buildings & 9.3 & 9 & 0.0\% \\

\end{xltabular}

\begin{xltabular}{\textwidth}{p{6cm}ccc}
   
    \caption{TURNOVER RATIOS - Controlled Scope} \label{tab10}\\
    \toprule
    Activities & Code & Turnover M\$ & Turnover Perc \\
    \midrule

    \endhead
    \bottomrule
    \endfoot

    Manufacture of low carbon technologies for transport & 3.3 & 370 & 0.1\% \\
    Manufacture of batteries & 3.4 & 833 & 0.3\% \\
    Manufacture of energy efficiency equipment for buildings & 3.5 & 29 & 0.0\% \\
    Manufacture of plastics in primary form & 3.17 & 47 & 0.0\% \\
    Electricity generation using solar photovoltaic technology & 4.1 & 45 & 0.0\% \\
    Electricity generation from wind power & 4.3 & 47 & 0.0\% \\
    Electricity generation from hydropower & 4.5 & 2 & 0.0\% \\
    Storage of electricity & 4.10 & 15 & 0.0\% \\
    Manufacture of biogas/biofuels for use in transport & 4.13 & 179 & 0.1\% \\
    District heating/cooling distribution & 4.15 & 73 & 0.0\% \\
    Production of heat/cool using waste heat & 4.25 & 1 & 0.0\% \\
    Anaerobic digestion of bio-waste & 5.7 & 32 & 0.0\% \\
    Landfill gas capture and utilisation & 5.10 & 8 & 0.0\% \\
    Infrast. enabling low-carbon road transport and public transport & 6.15 & 54 & 0.0\% \\
    Installation, maintenance and repair of charging stations for electric vehicles & 7.4 & 28 & 0.0\% \\
    Installation, maintenance and repair of renewable energy tech. & 7.6 & 1,398 & 0.5\% \\
    Professional services related to energy performance of buildings & 9.3 & 305 & 0.1\% \\

\end{xltabular}

\begin{xltabular}{\textwidth}{p{2.5cm} p{1cm} p{2.5cm} p{2cm} p{2.5cm}}
    \caption{Reasons and Explanations for Activities} \label{tab11} \\
    \toprule
    Activities & Link & Explanation & Contribution & Explanation \\
    \midrule
    \endhead
    \bottomrule
    \endfoot

    Afforestation & 0 & Does not directly reduce the need for oil extraction & 1 & Afforestation is a positive initiative that helps to reduce carbon dioxide emissions, but it does not directly reduce the need for oil extraction. \\
    Manufacture of low carbon technologies for transport & 4 & Can help to reduce the oil demand for transportation & 2 & The manufacture of low-carbon technologies, such as electric vehicles and hybrid vehicles, can help to reduce the oil demand for transportation. However, these technologies still require some oil for their production and operation. \\
    Manufacture of batteries & 3 & A key part of the supply chain for electric vehicles & 2 & The manufacture of batteries, which are used in electric vehicles and other low-carbon technologies, can also help to reduce the oil demand. However, the production of batteries requires some oil. \\
    Manufacture of energy efficiency equipment for buildings & 4 & Reduces the energy consumption of the buildings & 3 & The manufacture of energy-efficient equipment, such as insulation and windows, can help to reduce the energy consumption of buildings, which in turn reduces the oil demand for electricity generation and heating. \\
    Manufacture of organic basic chemicals & 3 & Invests in the production of renewable chemicals & 1 & The manufacture of organic basic chemicals, such as plastics and solvents, requires oil as a feedstock. Therefore, this activity does not directly contribute to the reducion of the need for oil extraction. \\
    Manufacture of plastics in primary form & 1 & Major consumer of oil and does not directly linked with Totalenergy's activity. & 0 & The manufacture of plastics in primary forms, such as polyethylene and polypropylene, is a major consumer of oil. Therefore, this activity directly contributes to increase oil demand. \\
    Electricity generation using solar photovoltaic technology & 3 & Does not require oil and contributes to a cleaner future & 3 & Electricity generation using solar photovoltaic (PV) technology does not require oil. Therefore, this activity directly reduces the oil demand and contributes to a cleaner energy future. \\
    Electricity generation from wind power & 4 & Does not require oil and contributes to a cleaner future & 3 & Electricity generation from wind power does not require oil. Therefore, this activity directly reduces the oil demand and contributes to a cleaner energy future. \\
    Electricity generation from hydropower & 3 & Does not require oil and contributes to a cleaner future & 3 & Electricity generation from hydropower does not require oil. Therefore, this activity directly reduces the oil demand oil and contributes to a cleaner energy future. Hydropower is a renewable source of electricity that can help to reduce the need for oil extraction. \\
    Storage of electricity & 3 & Helps to smooth out the variability of renewable energy & 3 & Electricity storage can help to smooth out the variability of renewable energy sources, such as solar and wind power. This can reduce the need for fossil fuels, including oil, for backup power generation. \\
    Manufacture of biogas/biofuels for use in transport & 3 & Reduces the oil demand for transportation & 2 & The manufacture of biogas/biofuels from renewable sources, such as biomass, can help to reduce the oil demand for transportation. However, this process can also release methane, a potent greenhouse gas. \\

    District heating/cooling distribution & 2 & Distributes heat from centralised sources to homes and businesses. Can help to reduce reliance on oil-fired heating and cooling systems. & 2 & District heating and cooling systems utilize heat from industrial processes or renewable sources, reducing the need for oil-based energy sources. \\
    
    Production of heat/cool using waste heat & 3 & Recovers waste heat from industrial processes & 2 & Production of heat/cool using waste heat from industrial processes or other sources can reduce the need for oil for energy consumption. \\
    Anaerobic digestion of bio-waste & 2 & Produces biogas, which can be used to generate electricity or heat & 2 & Anaerobic digestion of bio-waste, such as food scraps and yard waste, can produce biogas, which can be used to generate electricity or heat. This can help to reduce the oil demand. \\

    Landfill gas capture and utilization & 3 & Recovers methane from landfills, which can be used to generate electricity or heat. Methane is a potent greenhouse gas. & 2 & Landfill gas capture and utilization can reduce the amount of methane released into the atmosphere, mitigating its impact on climate change. \\
     
    Underground permanent geological storage of CO2 & 4 & Mitigates the climate impact of oil production & 1 & Underground permanent geological storage of CO2 can help to mitigate the climate impact of oil production. However, it is a complex and expensive process, and its effectiveness is still being debated. \\
    Infrast. enabling low-carbon road transport and public transport & 3 & Supports the deployment of low-carbon transport technologies & 2 & Infrastructure that enables low-carbon road transport, such as charging stations for electric vehicles, can help to promote the adoption of electric vehicles and reduce the oil demand. \\
    Construction of new buildings & 3 & Can be designed to be energy efficient & 1 & The construction of new buildings can contribute to the oil demand, as it requires materials that are often produced using oil-based products. \\
    Installation, maintenance and repair of charging stations for electric vehicles & 3 & Supports the adoption of electric vehicles & 2 & Installation, maintenance, and repair of charging stations for electric vehicles can help to ensure their availability and reliability, which can encourage the adoption of electric vehicles and reduce the oil demand. \\
    Installation, maintenance and repair of renewable energy tech. & 2 & Ensures the continued operation and efficiency of renewable energy systems & 3 & Installation, maintenance, and repair of renewable energy technologies, such as solar panels and wind turbines, can help to ensure their continued operation and efficiency, which can further reduce the oil demand. \\
    Acquisition and ownership of buildings & 1 & Can be used to acquire and operate energy-efficient buildings & 1 & Acquisition and ownership of buildings can contribute to the oil demand, as they require energy for heating, cooling, and lighting. \\
    Professional services related to energy performance of buildings & 1 & Provides guidance and expertise on energy efficiency & 3 & Professional services related to the energy performance of buildings, such as energy audits and energy efficiency upgrades, can help to reduce the energy consumption of buildings and thereby reduce the oil demand for electricity generation and heating.
\end{xltabular}


\newpage

\bibliographystyle{plainnat}
\bibliography{bibliography} 
\newpage
\end{document}